\documentstyle[eqsecnum,aps]{revtex}

\begin{document}
\draft
\preprint{HEP/123-qed}
\title{Asymptotes and characteristic times for transmission and
reflection.}

\author{N.L.Chuprikov} \address{Tomsk State Pedagogical
University \\ 634041 Tomsk, Russia} \date{\today}

\maketitle

\begin{abstract}

A complete one-dimensional scattering of a spinless particle on a
time-independent potential barrier is considered. To describe
separately transmitted and reflected particles in the corresponding
subsets of identical experiments, we introduce the notions of
scattering channels for transmission and reflection. We find for both
channels the (unitary) scattering matrices and reconstruct, by known
out asymptotes (i.e., by the transmitted and reflected wave packets),
the corresponding in asymptotes. Unlike the out asymptotes for
transmission and reflection, their in asymptotes represent
nonorthogonal functions. As is shown, the position distributions of
to-be-transmitted and to-be-reflected particles, except their average
positions, are unpredictable. At the same time, the momentum
distributions of these particles are physically meaningful and can be
observed to the full. We show that both the subensembles of particles
must start (on the average) from the same spatial point, and the
momentum distributions of to-be-scattered and scattered particles must
be the same for each scattering channel.  Taking into account these
properties, we define the (individual) delay times for transmission and
reflection for wave packets of an arbitrary width.  Besides, to
estimate the duration of the scattering event, we derive the expression
for a (total) scattering time.

\end{abstract}
\pacs{PACS numbers:  03.65.Ca,03.65.Xp }
\twocolumn

\newcommand{\Api}{A^{(+)}_{in}}
\newcommand{\Ami}{A^{(-)}_{in}}
\newcommand{\Apo}{A^{(+)}_{out}}
\newcommand{\Amo}{A^{(-)}_{out}}

\section*{Introduction}

For a long time tunneling a spinless particle through a one-dimensional
static potential barrier was considered in quantum mechanics as a
representative of well-understood phenomena. However, now it has been
realized that this is not the case. The inherent to quantum theory
standard wave-packet analysis (SWPA) (see, for example,
\cite{Ha2,Ha1,Ter}) does not provide a clear prescription both how to
interpret properly the temporal behavior of finite in $x$ space wave
packets and how to introduce characteristic times for a tunneling
particle.

As is known, the main peculiarity of the tunneling of finite wave
packets is that the average particle's momentum for the transmitted,
reflected and incident wave packets are different. It is evident
that this fact needs in a proper explanation. As was pointed out in
\cite{La1}, it would be strange to interpret the above property of wave
packets as the evidence of accelerating a particle (in the asymptotic
regions) by a static potential barrier. Besides, in this case there
is no causal link between the transmitted and incident wave packets
(see \cite{La1}), and the tunneling times introduced in the SWPA are
ill-defined \cite{Ha2,La1}.

Perhaps, no problems in this approach arise only for wide (strictly
speaking, infinite) in $x$ space wave packets, when the average kinetic
energy of particles, before and after the interaction, is the same. In
this case the asymptotic phase times \cite{Ha2,Ha1,Ter,Col} are
formally well-defined.

Apart from the SWPA, the variety of alternative approaches to solving
the tunneling time problem (TTP) have also been developed (see, for
example,
\cite{Ha2,Ha1,Ter,La1,Mu0,Col,Ja1,Nus,Mu6,Aha,Pau,La2,Sok,Nas,Le1,La3,Olk,Mu4,Ste,Yam}
and references therein). However again, a clear agreement between
these approaches and the SWPA have been achieved only for wide wave
packets.  In the general case all the attempts have not yet led to
commonly accepted quantities which would describe a tunneling
particle in the standard setting the tunneling problem
\cite{Ha2,La1,Mu0}. As regards the interpretation of the above
peculiarity of the tunneling of finite wave packets, as we are aware
this question was of no interest in these approaches.

However, we think the question of timing a tunneling particle cannot be
solved without of a proper understanding of the behavior of tunneling
wave packets. In this paper we develop the approach, in which both the
questions are closely connected. Its basis is the formalism of a
separate description of transmitted and reflected particles at the
stage preceding the scattering event. We show that such a description
is needed and quite admissible in quantum theory.

The necessity in the separate description follows from the fact that
when $t\to \infty$ the transmitted and reflected wave packets are
localized, in the case of a completed scattering, in the disjoint
spatial regions. By the ensemble (statistical) interpretation of
quantum mechanics, this means that the whole (infinite) set of
identical experiments on tunneling, in which scattered particles are
detected far from the barrier, is divided into two separate parts. One
of them includes experiments in which a particle is transmitted through
the barrier: we will say that it moves in this case along the
transmission channel. In another part, a particle is reflected by the
barrier: similarly, we will say that it moves in such experiments along
the reflection channel.

In each experiment a particle is evident to pass all the distance
between the particle's source and detector. So that at the initial
stage of scattering we deal in fact with the subensembles of
to-be-transmitted (STP) and to-be-reflected (SRP) particles. Of course,
the formalism of quantum mechanics, as it stands, does not suggest a
separate description of the STP and SRP. The states of these particles
cannot be described in terms of orthogonal wave functions. However, it
is evident that just the STP and SRT should be causally connected with
the subensembles of transmitted and reflected particles, respectively.
Thus, to clear up the above acceleration of finite wave packets, one
needs to study the dynamics of these subensembles both after and before
the scattering event.

As will be shown in this paper, the principles of quantum mechanics
admit a separate description of the STP and SRP. One can uniquely
introduce the non-orthogonal in asymptotes for transmission and
reflection.  Being "unphysical" from the viewpoint of the conventional
quantum description, they in reality possess some properties typical for
"physical" (orthogonal) quantum states. In particular, the number of
particles in each subensemble should be conserved. The momentum
distributions of to-be-transmitted and to-be-reflected particles are
physically meaningful and should be experimentally observed. In
particular, the energy distribution of particles in the asymptotic
regions, to the left and to the right of the barrier, should be the
same for each scattering channel. At the same time, due to the
statistical dependency of these subensembles, their position
distributions cannot be, in full, reproducible in experiment. Only the
average positions of to-be-transmitted and to-be-reflected particles
can be uniquely determined and verified experimentally.

Taking into account the properties of the in asymptotes for
transmission and reflection, we offer our interpretation to the
behavior of finite wave packets in tunneling and define delay times for
transmission and reflection.  Besides, apart from the separate
description of transmitted and reflected particles, we introduce a
total scattering time to estimate the duration of the scattering event.

The paper is organized as follows. In Section \ref{a1} we find total in
and out asymptotes for a tunneling particle, and display explicitly
shortcomings to arise in the SWPA in solving the TTP. In Section
\ref{a3} we introduce the notion of in asymptotes for transmission and
reflection and define the corresponding (individual) delay times which
can be verified experimentally. The (total) scattering time to describe
the duration of the scattering event, as well as its start and finish
instants of time are derived in Section \ref{a5}. In addition, we find
here the restrictions on the wave-packet's parameters which must be
fulfilled for a completed scattering.  The expressions for the
asymptotic expectation values of the position and wave-number operators
as well as for their mean-square deviations are presented in Appendix.

\newcommand{\ko}{\kappa_0^2}
\newcommand{\kj}{\kappa_j^2}
\newcommand{\kd}{\kappa_j d_j}
\newcommand{\kki}{\kappa_0\kappa_j}

\newcommand{\Ra}{R_{j+1}}
\newcommand{\Rb}{R_{(1,j)}}
\newcommand{\Rc}{R_{(1,j+1)}}

\newcommand{\Ta}{T_{j+1}}
\newcommand{\Tb}{T_{(1,j)}}
\newcommand{\Tc}{T_{(1,j+1)}}

\newcommand{\Wa}{w_{j+1}}
\newcommand{\Wb}{w_{(1,j)}}
\newcommand{\Wc}{w_{(1,j+1)}}

\newcommand{\UU}{u^{(+)}_{(1,j)}}
\newcommand{\VV}{u^{(-)}_{(1,j)}}

\newcommand{\ta}{t_{j+1}}
\newcommand{\tb}{t_{(1,j)}}
\newcommand{\tc}{t_{(1,j+1)}}

\newcommand{\tee}{\vartheta_{(1,j)}}

\newcommand{\tta}{\tau_{j+1}}
\newcommand{\ttb}{\tau_{(1,j)}}
\newcommand{\ttc}{\tau_{(1,j+1)}}

\newcommand{\FF}{\chi_{(1,j)}}
\newcommand {\aro}{(k)}

\section{Setting the problem for a completed scattering}
\label{a1}
\subsection{Backgrounds} \label{a10}

Let us consider a particle that tunnels through the time-independent
potential barrier $V(x)$ confined to the finite spatial interval
$[a,b]$ $(a>0)$; $d=b-a$ is the barrier width. Let its in state,
$\Psi_{in}(x),$ at $t=0$ be the normalized function
$\Psi^{(0)}_{left}(x)$ when it moves from the left, or
$\Psi^{(0)}_{right}(x)$, otherwise. Both functions belongs to the set
$S_{\infty}$ to consist from infinitely differentiable functions
vanishing exponentially in the limit $|x|\to \infty$. The
Fourier-transforms of such functions are known to belong to the set
$S_{\infty}$, too. In this case the position and momentum operators both
are well-defined. Without loss of generality we will suppose that

\[<\Psi^{(0)}_{left}|\hat{x}|\Psi^{(0)}_{left}>=0, \hspace{3mm}
<\Psi^{(0)}_{left}|\hat{p}|\Psi^{(0)}_{left}> =\hbar k_0 > 0,\]
\[<\Psi^{(0)}_{left}|\hat{x}^2|\Psi^{(0)}_{left}> =l_0^2,
\hspace{3mm}
\Psi^{(0)}_{right}(x)\equiv \Psi^{(0)}_{left}(x_r-x)\]
(that is, the function $\Psi^{(0)}_{right}(x)$ is centered at the
point $x=x_r$); here $l_0$ is the wave-packet's half-width at $t=0$
($l_0<<a$ and $l_0<<x_r-b$); $\hat{x}$ and $\hat{p}$ are the operators
of the particle's position and momentum, respectively.

Besides, let $\hat{H}$ be the Hamiltonian, $\hat{H}=\hat{H}_0+V(x)$
where $\hat{H}_0$ describes a free particle. Let also $\hat{H}_0^{ref}$
be the Hamiltonian to describe the ideal reflection of a free particle
off the potential wall located at the middle point of the interval
$[a,b]$, i.e., at $x_{midp}=(a+b)/2$.

An important restriction should be imposed on the rate of spreading
the incident wave packet. We must be sure that at early times the
quantum ensemble of particles moves, as a whole, toward the barrier. In
particular, at the initial stage of scattering, the probability to find
a particle at the initial point should diminish in time. This does not
at all mean that the incident wave packet must not contain waves with
$p\le 0$ ($p\ge 0$), if the initial state of a particle is described by
the function $\Psi^{(0)}_{left}$ ($\Psi^{(0)}_{right}$). This condition
is fulfilled when the back front of the incident wave-packet, which is
away from the center of mass (CM) of the wave packet at the distance
equal to the wave-packet's half-width, moves toward the barrier. Such a
behavior takes place only if the packet's spreading is ineffective
enough (see condition (\ref{304}) in Section \ref{a5}).

\subsection{Stationary states} \label{a11}

As is known, the formal solution to the temporal one-dimensional
Schr\"odinger equation (OSE) of the problem at hand can be written as
$e^{-i\hat{H}t/\hbar}\Psi_{in}(x).$ To solve explicitly this equation,
we will use here the transfer matrix method (TMM) \cite{Ch1} that
allows one to calculate the tunneling parameters for any system of
potential barriers. The state of a particle with the wave-number $k$
can be written in the form

\begin{eqnarray} \label{1}
\Psi_{left}=\left[\Api(k)\exp(ikx)
+\Amo(k)\exp(-ikx)\right] \nonumber \\ \times\exp[-iE(k)t/\hbar],
\end{eqnarray}
for $x\le a$; and for $x>b$ we have

\begin{eqnarray} \label{2}
\Psi_{right}=\left[\Apo(k)\exp(ikx)
+\Ami(k)\exp(-ikx)\right] \nonumber \\ \times\exp[-iE(k)t/\hbar].
\end{eqnarray}

\noindent Here $E(k)=\hbar^2 k^2/2m.$ The coefficients entering this
solution are connected by the transfer matrix ${\bf Y}$ (see
\cite{Ch1}):

\begin{eqnarray} \label{50}
\left(\begin{array}{c} \Api \\ \Amo
\end{array} \right)={\bf Y} \left(\begin{array}{c} \Apo \\ \Ami
\end{array} \right); \hspace{8mm}
{\bf Y}=\left(\begin{array}{cc} q & p \\ p^* &
q^* \end{array} \right);
\end{eqnarray}
where

\[
q=\frac{1}{\sqrt{T(k)}}\exp[-i(J(k)-kd)];
\]
\[
p=\sqrt{\frac{R(k)}{T(k)}}\exp[i(\frac{\pi}{2}+F(k)-ks)];
\]

\noindent $T(k)$  (the real transmission coefficient) and $J(k)$
(phase) are even and odd functions of $k$, respectively;
$F(-k)=\pi-F(k)$; $R(k)=1-T(k)$; $s=a+b$. Note that the functions
$T(k)$, $J(k)$ and $F(k)$ contain all needed information about the
influence of the potential barrier on a particle. We will suppose that
these functions have already been known explicitly. To find them, one
can use the recurrence relations obtained in \cite{Ch1}.

The amplitudes of the outgoing and corresponding incoming waves are
connected by the scattering matrix ${\bf S}$:

\begin{eqnarray} \label{51}
{\cal A}_{out}= {\bf S} {\cal A}_{in}; \hspace{3mm}
{\bf S}= \left(\begin{array}{cc} S_{11} & S_{12} \\ S_{21} & S_{22}
\end{array} \right);
\end{eqnarray}

\[{\cal A}_{in}=\left(\begin{array}{c} \Api \\ \Ami
\end{array} \right); \hspace{3mm} {\cal A}_{out}= \left(\begin{array}{c}
\Apo \\ \Amo \end{array} \right);\]
here
\begin{eqnarray} \label{52}
S_{11}=S_{22}=q^{-1}=\sqrt{T(k)}\exp[i(J-kd)],\nonumber \\
S_{12}=-\frac{p}{q}=\sqrt{R(k)}\exp[i(J+F-\frac{\pi}{2}-2kb)]
\nonumber \\
S_{21}=\frac{p^*}{q}=\sqrt{R(k)}\exp[i(J-F-\frac{\pi}{2}+2ka)].
\end{eqnarray}

Note that the scattering matrix can be uniquely presented as the sum of
two non-unitary matrices to describe separately transmission and
reflection,

\begin{eqnarray} \label{520}
{\bf S}={\bf \Pi}_{tr}+ {\bf \Pi}_{ref}
\end{eqnarray}
where
\begin{eqnarray} \label{53}
{\bf \Pi}_{tr}={\bf S}_{tr} {\bf P}_{tr},
\hspace{3mm}
{\bf \Pi}_{ref}={\bf S}_{ref} {\bf P}_{ref}; \nonumber\\
{\bf S}_{tr}={\bf \Delta}_{tr} {\bf S}^{(0)}_{tr}, \hspace{3mm}
{\bf S}_{ref}={\bf \Delta}_{ref} {\bf S}^{(0)}_{ref}
\end{eqnarray}
(${\bf S}_{tr}$ and ${\bf S}_{ref}$ are unitary matrices);
\begin{eqnarray} \label{54}
{\bf P}_{tr}={\bf I} \sqrt{T}; \hspace{3mm}
{\bf P}_{ref}={\bf I} \sqrt{R} ;
\end{eqnarray}
\begin{eqnarray} \label{55}
{\bf S}^{(0)}_{tr}={\bf I};
\hspace{3mm}
{\bf S}^{(0)}_{ref} = \left(\begin{array}{cc} 0 & -e^{-iks} \\
-e^{iks} & 0 \end{array} \right);
\end{eqnarray}
\begin{eqnarray} \label{56}
{\bf \Delta}_{tr}={\bf I}\exp[i(J-kd)];\nonumber \\
{\bf \Delta}_{ref}=
\left(\begin{array}{cc} e^{iF} & 0 \\ 0 &
e^{-iF} \end{array}
\right)\exp[i(J+\frac{\pi}{2}-kd)];\end{eqnarray}
where ${\bf I}$ is the unit matrix; ${\bf S}^{(0)}_{tr}$ and ${\bf
S}^{(0)}_{ref}$ are the scattering matrix to correspond the
Hamiltonians $\hat{H}_0$ and $\hat{H}_0^{ref}$, respectively.

\subsection{Total in and out asymptotes} \label{a12}

As is known, solving the TTP is reduced in the SWPA to timing a
particle beyond the scattering region where the exact solution of the
OSE approaches to the corresponding in or out asymptote \cite{Tei}.
Thus, definitions of characteristic times for a tunneling particle can
be obtained in terms of the in and out asymptotes of the tunneling
problem. To find them, we have to go over to the temporal scattering
problem and consider two independent cases,
\begin{eqnarray} \label{88}
{\cal A}_{in}=\left(\begin{array}{c} \Api(k) \\ 0 \end{array}\right),
\hspace{3mm} {\cal A}_{in}=\left(\begin{array}{c} 0 \\ \Ami(k)
\end{array} \right),
\end{eqnarray}
when the incident particle moves toward the barrier from the left (the
left-side case) or from the right (the right-side case), respectively.

Note that in both the cases the in asymptotes represent one-packet
objects to converge, at $t\to -\infty,$ with the corresponding incident
packets, while the out asymptotes represent a superposition of two
non-overlapped wave packets to converge, at $t\to \infty,$ with the
superposition of the transmitted and reflected ones. It is easy to show
that in the first case the in and out asymptotes, in $k$ space, can be
written for both scattering channels as follows

\begin{eqnarray} \label{59}
f_{in}(k,t)=\Api(k) \exp[-i E(k)t/\hbar];
\end{eqnarray}

\begin{eqnarray} \label{60}
f_{out}(k,t)= f_{out}^{tr}(k,t)+f_{out}^{ref}(k,t)
\end{eqnarray}
where
\begin{eqnarray} \label{61}
f_{out}^{tr}(k,t)=\sqrt{T(k)}\Api(k) \exp[i(J(k)-kd \nonumber
\\-E(k)t/\hbar)];
\end{eqnarray}

\begin{eqnarray} \label{62}
f_{out}^{ref}(k,t)=\sqrt{R(k)}\Api(-k)
\exp[-i(J(k) \nonumber \\-F(k)-\frac{\pi}{2}+2ka+E(k)t/\hbar)].
\end{eqnarray}

For particles starting, on the average, at the origin (the left-side
case), we have (see Appendix)
\begin{eqnarray} \label{63}
<\hat{x}>_{in}=\frac{\hbar k_0}{m}t.
\end{eqnarray}
The averaging separately over the transmitted and reflected wave
packets yields

\begin{eqnarray} \label{64}
<\hat{x}>^{tr}_{out}=\frac{\hbar t}{m}<k>^{tr}_{out}
-<J^\prime(k)>^{tr}_{out}+d;
\end{eqnarray}
\begin{eqnarray} \label{65}
<\hat{x}>^{ref}_{out}=\frac{\hbar t}{m}<k>^{ref}_{out}
 \nonumber \\+<J^\prime(k)-F^\prime(k)>^{ref}_{out}+2a
\end{eqnarray}
(hereinafter the prime denotes the derivative with respect to $k$).

Similarly, for the right-side case, we have
\begin{eqnarray} \label{67}
f_{in}(k,t)=\Ami(-k) \exp[-i E(k)t/\hbar];
\end{eqnarray}
\begin{eqnarray} \label{68}
f_{out}^{tr}(k,t)=\sqrt{T(k)}\Ami(-k)
\exp[-i(J(k)-kd \nonumber \\+E(k)t/\hbar)];
\end{eqnarray}
\begin{eqnarray} \label{700}
f_{out}^{ref}(k,t)=\sqrt{R(k)}\Ami(k)
\exp[i(J(k) \nonumber \\+F(k)-\frac{\pi}{2}-2kb-E(k)t/\hbar)].
\end{eqnarray}
Hence
\begin{eqnarray} \label{710}
<\hat{x}>_{in}=x_r-\frac{\hbar k_0}{m}t;
\end{eqnarray}
\begin{eqnarray} \label{720}
<\hat{x}>^{tr}_{out}=x_r+\frac{\hbar t}{m}<k>^{tr}_{out}
+<J^\prime(k)>^{tr}_{out}-d;
\end{eqnarray}
\begin{eqnarray} \label{74}
<\hat{x}>^{ref}_{out}=-x_r+\frac{\hbar t}{m}<k>^{ref}_{out}
 \nonumber \\-<J^\prime(k)+F^\prime(k)>^{ref}_{out}+2b.
\end{eqnarray}

\subsection {Paradoxes of the standard wave-packet analysis} \label{a2}

To display explicitly the basic shortcoming of the SWPA, let us derive
again the SWPA's tunneling times. For this purpose it is sufficient to
consider the left-side case.

Let $Z_1$ be a point to lie at some distance $L_1$ ($L_1\gg l_0$ and
$a-L_1\gg l_0$) from the left boundary of the barrier, and $Z_2$ be a
point to lie at some distance $L_2$ ($L_2\gg l_0$) from its right
boundary. Following \cite{Ha1}, let us define the difference between
the times of arrival of the CMs of the incident and
transmitted packets at the points $Z_1$ and $Z_2$, respectively (this
time will be called below as the "transmission time").  Analogously,
let the "reflection time" be the difference between the times of
arrival of the CMs of the incident and reflected packets at the same
point $Z_1$.

\newcommand{\ppp}{\mbox{\hspace{5mm}}}
\newcommand{\ooo}{\mbox{\hspace{3mm}}}

Thus, let $t_1$ and $t_2$ be such instants of time that

\begin{equation} \label{8}
<\hat{x}>_{in}(t_1)=a-L_1; \ooo
<\hat{x}>^{tr}_{out}(t_2)=b+L_2.
\end{equation}

\noindent Then, considering (\ref{63}) and (\ref{64}), one can write the
"transmission time" $\Delta t_{tr}$ ($\Delta t_{tr} =t_2 -t_1$) for the
given interval in the form

\begin{eqnarray} \label{9}
\Delta t_{tr}=\frac{m}{\hbar}\Big[\frac{<J^\prime>_{out}^{tr} +L_2}
{<k>_{out}^{tr}} +\frac{L_1}{k_0} \nonumber \\ +
a\left(\frac{1}{<k>_{out}^{tr}} -\frac{1}{k_0}\right)\Big].
\end{eqnarray}
Similarly, for the reflected packet, let $t^{\prime}_1$ and
$t^{\prime}_2$ be such instants of time that

\begin{equation} \label{10}
<\hat{x}>_{in}(t^{\prime}_1)
=<\hat{x}>_{out}^{ref}(t^{\prime}_2)=a-L_1.
\end{equation}

\noindent From equations (\ref{63}), (\ref{65}) and (\ref{10}) it
follows that the "reflection time" $\Delta t_{ref}$ ($\Delta
t_{ref}=t^{\prime}_2-t^{\prime}_1$) can be written as

\begin{eqnarray} \label{11}
\Delta t_{ref}=\frac{m}{\hbar}\Big[\frac{<J^\prime -
F^\prime>_{out}^{ref} +L_1} {<-k>_{out}^{ref}} +\frac{L_1}{k_0}
\nonumber\\ +a\left(\frac{1}{<-k>_{out}^{ref}}
-\frac{1}{k_0}\right)\Big].
\end{eqnarray}

Note that the expectation values of $k$ for all three wave packets
coincide only in the limit $l_0\to\infty$ (i.e., for particles with a
well-defined momentum). In the general case these quantities are
distinguished. For example, for a particle whose initial state, in the
left-side case (\ref{88}), is described by the Gaussian wave packet
(GWP),
\[\Api(k)=\frac{l_0}{\sqrt{\pi}}\exp(-l_0^2(k-k_0)^2),\]
we have

\begin{equation} \label{100}
<k>_{tr}=k_0+\frac{<T^\prime>_{in}} {4l_0^2<T>_{in}};
\end{equation}

\begin{equation} \label{101}
<-k>_{ref}=k_0+\frac{<R^\prime>_{in}} {4l_0^2<R>_{in}}.
\end{equation}
Let
\[<k>_{tr}=k_0+(\Delta k)_{tr},\ppp <-k>_{ref}=k_0+(\Delta k)_{ref},\]

\noindent then relations (\ref{100}) and (\ref{101}) can be written in
the form

\begin{equation} \label{102}
\bar{T}\cdot (\Delta k)_{tr}=-\bar{R}\cdot(\Delta k)_{ref}
=\frac{<T^\prime>_{in}}{4l_0^2}.
\end{equation}
\noindent Note that $R^\prime=-T^\prime$.

As is seen, in the general case quantities (\ref{9}) and (\ref{11})
cannot serve as characteristic times for a particle.  Due to the last
terms in these expressions the above times depend essentially on the
initial distance between the wave packet and barrier, with $L_1$ being
fixed.  These terms are dominant for the sufficiently large distance
$a$.  Moreover, one of them must be negative. For example, for the
transmitted wave packet it takes place in the case of the under-barrier
tunneling through an opaque rectangular barrier. The numerical
modelling of tunneling \cite{Ha2,Ha1,Ter,Le1} shows in this case a
premature appearance of the CM of the transmitted packet behind the
barrier, what points to the lack of a causal link between the
transmitted and incident wave packets (see \cite{La1}).

As was shown in \cite{Ha2,Ha1}, this effect disappears in the limiting
case $l_0\to\infty$. In the case of Gaussian wave packets, the fact
that the last terms in (\ref{9}) and (\ref{11}) tend to zero when
$l_0\to \infty$, with the ratio $l_0/a$ being fixed, can be proved with
help of Exps. (\ref{100}) and (\ref{101}) (note that the limit $l_0\to
\infty$ with a fixed value of $a$ is unacceptable in this analysis,
because it contradicts the initial condition $a\gg l_0$ for a completed
scattering). Thus, in the limit $l_0\to\infty$ the SWPA formally
provides characteristic times for a particle.

Note, the fact that Exps. (\ref{100}) and (\ref{101}) cannot be
applied to particles does not at all mean that they are erroneous.
These expressions correctly describe the relative motion of the
transmitted (or reflected) and incident wave packets. The main problem
is to understand what particle's dynamics underlies such a behavior of
wave packets.

\section {Formalism of separate description of transmitted and
reflected particles} \label{a3}

\subsection {In and out asymptotes for transmission and reflection}
\label{a31}

\hspace*{\parindent} We think that the principle mistake made in the
SWPA in deriving the individual tunneling times is that the incident
wave packet cannot be used as a counterpart neither to the transmitted
nor to reflected wave packet, when they are treated separately. This
step is physically meaningless because the incident packet describes
the whole quantum ensemble of particles (or, the whole set of the
corresponding identical experiments), while the transmitted packet, for
example, represents only its part. The former can be used as a
reference only for the transmitted and reflected packets taken jointly.
As regards the separate description of transmitted particles, for
example, it makes sense to compare their motion only with that of
to-be-transmitted particles mentioned in Introduction. This means that
in order to develop a separate description of both scattering channels,
one needs to find the in and out asymptotes for transmission and
reflection, if they exist.

Let us show that decomposition (\ref{520}) permits us to find uniquely
such asymptotes. Indeed, let

\[{\cal A}_{out}^{tr}= {\bf \Pi}_{tr} {\cal A}_{in}; \hspace{3mm}
{\cal A}_{out}^{ref}= {\bf \Pi}_{ref} {\cal A}_{in}.\]
Considering (\ref{53}) we can rewrite these relations as follows,
\begin{eqnarray} \label{57}
{\cal A}_{out}^{tr}={\bf S}_{tr} {\cal
A}_{in}^{tr}; \ppp {\cal A}_{out}^{ref}={\bf
S}_{ref} {\cal A}_{in}^{ref},
\end{eqnarray}
where
\begin{eqnarray} \label{58}
{\cal A}_{in}^{tr}= {\bf P}_{tr} {\cal A}_{in}; \ppp
{\cal A}_{in}^{ref}= {\bf P}_{ref} {\cal A}_{in}.
\end{eqnarray}
The pairs $({\cal A}_{in}^{tr},{\cal A}_{out}^{tr})$ and $({\cal
A}_{in}^{ref},{\cal A}_{out}^{ref})$ related by the unitary matrices
${\bf S}_{tr}$ and ${\bf S}_{ref}$, respectively, will be treated
further as the amplitudes of incoming and outgoing waves to describe
transmission and reflection. Since the amplitudes of outgoing waves are
known, relations (\ref{57}) can be used for reconstructing those of
incoming waves,
\begin{eqnarray*}
{\cal A}_{in}^{tr}={\bf S}^{-1}_{tr} {\cal A}_{out}^{tr}; \ppp
{\cal A}_{in}^{ref}={\bf S}^{-1}_{ref} {\cal A}_{out}^{ref}.
\end{eqnarray*}

Then it is easy to show that in the left-side case the searched-for in
asymptotes can be written, in $k$ space, as follows

\begin{eqnarray} \label{590}
f_{in}^{tr}(k,t)=\sqrt{T(k)}\Api(k) \exp[-i E(k)t/\hbar];
\end{eqnarray}
\begin{eqnarray} \label{610}
f_{in}^{ref}(k,t)=\sqrt{R(k)}\Api(k) \exp[-i E(k)t/\hbar];
\end{eqnarray}

Thus, for the left-side case (see (\ref{88}),
\begin{eqnarray} \label{630}
<\hat{x}>^{tr}_{in}=\frac{\hbar t}{m}<k>^{tr}_{in};
\end{eqnarray}
\begin{eqnarray} \label{650}
<\hat{x}>^{ref}_{in}=\frac{\hbar t}{m}<k>^{ref}_{in};
\end{eqnarray}

Similarly, for the right-side case, we have
\begin{eqnarray} \label{670}
f_{in}^{tr}(k,t)=\sqrt{T(k)}\Ami(-k) \exp[-i E(k)t/\hbar];
\end{eqnarray}
\begin{eqnarray} \label{690}
f_{in}^{ref}(k,t)=\sqrt{R(k)}\Ami(-k) \exp[-i E(k)t/\hbar].
\end{eqnarray}
As a consequence,
\begin{eqnarray} \label{711}
<\hat{x}>^{tr}_{in}=x_r+\frac{\hbar t}{m}<k>^{tr}_{in};
\end{eqnarray}
\begin{eqnarray} \label{730}
<\hat{x}>^{ref}_{in}=x_r+\frac{\hbar t}{m}<k>^{ref}_{in}.
\end{eqnarray}

Note that the separate treating of the out asymptotes for transmission
and reflection have been of usual practice. They are the transmitted
and reflected wave packets that coincide, in the limit $t\to \infty,$
with these asymptotes. For these orthogonal states which describe
mutually exclusive events, the probabilities $\bar{T}$ and $\bar{R}$
satisfy relation (\ref{2100}):  $\bar{T}+\bar{R}=1$.  Besides,
according to (\ref{212}),
\[<k^n>_{out}=\bar{T}<k^n>_{out}^{tr}+\bar{R}<(-k)^n>_{out}^{ref}.\]

It is important to emphasize that similar probabilistic rules take
place also for $f_{in}^{tr}(k,t)$ and $f_{in}^{ref}(k,t)$ to evolve
along the in asymptotes for transmission and reflection, respectively:

\begin{eqnarray} \label{901}
<f_{in}|f_{in}> =<f_{in}^{tr}|f_{in}^{tr}> +
<f_{in}^{ref}|f_{in}^{ref}>,
\end{eqnarray}

\begin{eqnarray} \label{902}
<k^n>_{in}=\bar{T}<k^n>_{in}^{tr}+\bar{R}<k^n>_{in}^{ref},
\end{eqnarray}

\begin{eqnarray} \label{903}
<\hat{x}>_{in}=\bar{T}<\hat{x}>_{in}^{tr}+
\bar{R}<\hat{x}>_{in}^{ref}.
\end{eqnarray}

However, because of the nonorthogonality of $f_{in}^{tr}$ and
$f_{in}^{ref},$ there are no similar probabilistic rules for higher
moments of the operator $\hat{x}$.

For each scattering channel we have
\begin{eqnarray} \label{904}
<f_{in}^{tr}|f_{in}^{tr}> =<f_{out}^{tr}|f_{out}^{tr}> =\bar{T}
\end{eqnarray}
and
\begin{eqnarray} \label{905}
<f_{in}^{ref}|f_{in}^{ref}> =<f_{out}^{ref}|f_{out}^{ref}> =\bar{R};
\end{eqnarray}
that is, the number of particles in each subensemble is the same
before and after the scattering event. Besides,

\begin{eqnarray} \label{906}
<\hat{k}>^{tr}_{in} =<\hat{k}>^{tr}_{out}
\end{eqnarray}
and

\begin{eqnarray} \label{907}
<\hat{k}>^{ref}_{in} =-<\hat{k}>^{ref}_{out};
\end{eqnarray}
which point to the conservation of the average momentum of transmitted
and reflected particles in the asymptotic spatial regions.

\subsection {Ideal and nonideal passage of particles in the
scattering channels} \label{a8}

So, in the tunneling problem considered as a two-channel scattering,
the transmission and reflection channels (or, what is equivalent, two
subsets of identical experiments, each includes only transmitted or
reflected particles) are described by Exps.  (\ref{901}) - (\ref{907}).
A simple analysis shows that it is convenient to distinguish a nonideal
and ideal passage of particles along the scattering channels. The ideal
passage in the transmission channel is characterized by the Hamiltonian
$\hat{H}_0$ and by the unit scattering matrix ${\bf S}^{(0)}_{tr}$
describing a free particle. By the ideal reflection is meant the
reflection of a free particle off the absolutely opaque potential wall
located at the point $x_{midp}$, which is described by the Hamiltonian
$\hat{H}_0^{ref}$ and scattering matrix ${\bf S}^{(0)}_{ref}$ (see
(\ref{55})).

In the general case, the scattering channels are always nonideal. A
nonideal passage of a particle along transmission and reflection
channels is described by the unitary matrices ${\bf S}_{tr}$ and ${\bf
S}_{ref}$, respectively.  As is seen from (\ref{53}) and (\ref{56}), in
this case a particle stay longer in the scattering region than in the
case of ideal scattering. In the subset of experiments where particles
are transmitted, the influence of the potential barrier on a particle
is equivalent, in the asymptotic regions, to that of a reflectionless
potential. In another subset its effect is similar to that of some
potential structure with the above opaque wall.

As it follows from the given formalism, the $\bar{T}$-th part of
particles moves in the transmission channel, while the $\bar{R}$-th
part does in the reflection one. The corresponding in and out
asymptotes are presented by expressions (\ref{61}), (\ref{62}),
(\ref{590}) and (\ref{610}) for the left-side case, or (\ref{68}),
(\ref{700}), (\ref{670}) and (\ref{690}) for the right-side case. The
particular cases, when $R(k)\equiv 0$ or $T(k)\equiv 0$, correspond to
the one-channel scattering processes: the transmission of a particle
through a reflectionless potential, and the reflection of a particle
off a structure containing the absolutely opaque potential wall.

The main result of the above formalism is that, for this elastic
quantum scattering, it extends the application of the conservation law
for the particle's momentum in the asymptotic regions to each
scattering channel. This automatically provides the momentum
distributions for to-be-transmitted and to-be-reflected particles.
Another result, consisting in that the average starting points for
to-be-transmitted and to-be-reflected particles are the same, reflects
eventually the fact that both the subensembles of particles are emitted
by the same source.

More detailed information about the individual properties of
transmitted and reflected particles at early times cannot be obtained
in quantum mechanics. For example, all the moments of the position
operator for the STP and SRP, excepting the first moment, cannot be
determined in principle. This implies that only the first moment of
the position operator may be used in defining individual characteristic
times for transmission and reflection.

\subsection {Delay times for transmitted and reflected particles}
\label{a4}

\hspace*{\parindent} Because of the influence of the potential barrier
a transmitted or reflected particle is delayed (on the average), in
the scattering region, relatively to a particle moving freely in the
scattering channel with the same in asymptote. For an observer
investigating only transmitted or reflected particles, it is important
to estimate the corresponding time and spatial delays. In particular,
by the sign of the time delay one can ascertain whether the potential
barrier investigated is repulsive or attractive with respect to the
given subensemble of particles.

\newcommand {\uta} {\tau _ {tr}}
\newcommand {\utb} {\tau _ {ref}}

At the beginning let us define the delay times for transmission and
reflection for the left-side case. As it follows from expressions
(\ref{64}) and (\ref{630}), the transmitted and corresponding free
particles arrive (on the average) at the same point $Z_2$ (see
\ref{a2}), at the instants $t^{tr}$ and $t^{tr}_{free}$, respectively,
such that

\begin{equation} \label{22}
<\hat{x}>_{in}^{tr}(t^{tr}_{free})=
<\hat{x}>_{out}^{tr}(t^{tr})= b+L_2.
\end{equation}

Then for transmitted particles the delay time $\tau^{tr}_{del}$ can be
defined as

\begin{eqnarray} \label{27}
\tau^{tr}_{del}=\frac{m} {\hbar<k>_{tr}}\left(<J^\prime>_{tr} -d
\right)
\end{eqnarray}
(since the average values of the tunneling parameters over the in and
out states are the same for both scattering channels, hereinafter we
will substitute $<\cdots>_{tr,ref}$  for $<\cdots>^{tr,ref}_{out}$).

Similarly, from expressions (\ref{65}) and (\ref{650}) it follows that
the reflected and corresponding ideally reflected particles arrive (on
the average) at the same point $Z_1$ at the instants $t^{ref}$ and
$t^{ref}_{free}$, respectively, such that

\begin{equation}
\label{23} <\hat{x}>_{in}^{ref}(t^{ref}_{free})=
<\hat{x}>_{out}^{ref}(t^{ref})= a-L_1.
\end{equation}

As a result, the delay time, $\tau^{(-)}_{del}$, for reflection can be
written in the form
\begin{eqnarray} \label{28}
\tau^{(-)}_{del}=\frac{m}{\hbar <-k>_{ref}}\bigg(<J^\prime -
F^\prime>_{ref} -d\bigg).
\end{eqnarray}
Analogously, for the right-side case the delay time for reflection is
given by

\begin{equation} \label{29}
\tau^{(+)}_{del}=\frac{m}{\hbar <k>_{ref}}\left(<J^\prime +
F^\prime>_{ref} -d\right)
\end{equation}
It is obvious that the transmission delay times for the left- and
right-side cases should be always equal. However, the equality
$\tau^{(+)}_{del}= \tau^{(-)}_{del}$ takes place only for symmetrical
potential barriers for which $F^\prime(k)\equiv 0$.

Note that the expressions $<J^\prime>_{tr} -d$ and $<J^\prime -
F^\prime>_{ref}-d$ (or $<J^\prime + F^\prime>_{ref} -d$) can be
treated as the spatial delays for the subensembles of transmitted and
reflected particles, respectively.

\subsection {About the verification of the individual properties of
to-be-transmitted and to-be-reflected particles.} \label{a9}

So, quantum theory quite admits a separate description of transmitted
and reflected particles at the stage preceding the scattering event. As
was shown, there can be uniquely determined, by known out asymptotes, in
asymptotes to describe individually to-be-transmitted and
to-be-reflected particles. The peculiarity of such description is that
these partial wave functions, being non-orthogonal, provide only such
physical characteristics as the momentum distributions, the average
positions and the delay times for to-be-transmitted and to-be-reflected
particles. Contrary to wave functions to describe the whole quantum
ensembles of free particles, the position distributions calculated over
its to-be-transmitted and to-be-reflected parts have, by our formalism,
no physical sense. These distributions should not be reproducible
(except the average positions), in the repeated series of identical
experiments.

Of course, a problem is that the dealing with the subensembles of
to-be-transmitted and to-be-reflected particles have not been inherent
in quantum mechanics. In particular, one may doubt in that the
particle subensembles described by non-orthogonal wave functions can
be, in principle, distinguishable.  Besides, by the above formalism the
reverse motion of the transmitted wave packet should describe only
particles which must pass through the barrier. However, this property
seems to contradict the rigorous results of one-dimensional quantum
theory. As is known, in the general case only a part of incident
particles may pass through the barrier.  Another part should be
reflected by it. So that there is the necessity to remove all these
doubts and to show that the above formalism is indeed in agreement
with the principles of quantum theory and can be experimentally
verified.

We will proceed from the natural assumption that in each single
experiment with a tunneling particle the latter interacts twice with an
experimental device: this takes place at the initial instant of time
when the particle is emitted in a given state by an appropriate source,
and at the final instant when it is absorbed by the detector to measure
the particle's final state. Note that to emit a free particle in the
given state means, in fact, to emit a particle with random
(unpredictable) values of the particle's momentum and position to
belong the given distributions. To be sure that a particle is indeed
emitted in the given state, in each single experiment an experimenter
should obtain (in the same or in the different series of identical
experiments with the particle) the values of the particle's momentum
and position not only for the final but also for the initial instant of
time. Due to time reversal, both the instants of time should be
equivalent in the sense that in the case of the reverse motion the
experimental data for the initial and final states should switch their
roles. Thus, from the above it follows that the division of
experimental data for scattered particles, in the corresponding
infinite set of identical experiments, should lead automatically to
that of data for to-be-scattered particles, thereby providing the
momentum and position distributions for to-be-transmitted and
to-be-reflected particles. So that, the subensembles of
to-be-transmitted and to-be-reflected are quite distinguishable.

To end this question, we have once more to pay reader's attention on the
status of the partial momentum and position distributions. The former
is reproducible, that is, one can carry out the several series of
identical experiments, and the momentum distributions for transmission
and reflection should be the same for these series. A cardinally
different situation should take place for their position distributions.
They are not reproducible, excepting the average positions of
to-be-transmitted and to-be-reflected particles.  The position
distributions for transmission and reflection obtained in two different
series of identical experiments may be different. For the given instant
of time, only the average positions of to-be-scattered particles for
each channel should be reproduced in this case.

Our next aim is to illustrate the difference between the cases when the
same wave packet, in one case, describes the whole ensemble of
particles, and in another case it does only a part. For this purpose it
is interesting to analyze the reverse motion of the transmitted and
reflected wave packets. Namely, let us consider three closely connected
solutions with the following in and out asymptotes:

1) the reverse motion of the transmitted wave packet -
\[f_{in}^{(1)}(k,t)= [f_{out}^{tr}(-k,t)]^*,\ooo f_{out}^{(1)}(k,t)=
[g_1(-k,t)]^*\]
where
\begin{eqnarray} \label{1001}
g_1(k,t)= \Big[T(k)\Api(k)-\sqrt{R(k)T(k)}\Api(-k)\nonumber \\
\times \exp[-i(\frac{\pi}{2}+F(-k)+ks)]\Big] \exp(-iE(k)t/\hbar);
\end{eqnarray}

2) the reverse motion of the reflected wave packet -
\[f_{in}^{(2)}(k,t)= [f_{out}^{ref}(-k,t)]^*,\ooo f_{out}^{(2)}(k,t)=
[g_2(-k,t)]^*\]
where
\begin{eqnarray} \label{1002}
g_2(k,t)= \Big[R(k)\Api(k) +\sqrt{R(k)T(k)}\Api(-k)\nonumber
\\ \times\exp[-i(\frac{\pi}{2}+F(-k)+ks)]\Big]\exp(-iE(k)t/\hbar);
\end{eqnarray}

3) the combined reverse motion -

\begin{eqnarray} \label{1003}
f_{in}^{(3)}(k,t)\equiv f_{in}^{(1)}(k,t) + f_{in}^{(2)}(k,t),
\nonumber\\ f_{out}^{(3)}(k,t) = [f_{in}(-k,t)]^*
\end{eqnarray}
(it is evident that the last asymptotes describe the motion that is
reverse with respect to the left-side case (see (\ref{88})).

As is seen, in the first (second) and third cases the incident wave
packets to the right (left) of the barrier are the same. This means
that in both the cases we deal with the same particle's source to the
right (left) of the barrier. It is evident that the momentum and
position distributions of emitted particles must be the same in these
cases, because the source's characteristics must not depend on the
availability of another source behind the barrier.

Now we have to take into account that in the first (second) case the
function $f_{in}^{(1)}$ ($f_{in}^{(2)}$) describes the whole set of
experiments but in the third case it does only a part. As is seen from
(\ref{1001}) and (\ref{1002}), the second terms in $f_{out}^{(3)}$ are
mutually cancelled. That is, the set of the experiments in the third
case does not contain experimental data to describe particles impinging
on the barrier from the right (and left) and then reflected by
(transmitted through) it. Thus, working with the right (left) source, in
the third case, the investigator has to take into account only those
experiments in which a particle passes through (is reflected by) the
barrier. By the above formalism, in this case the out asymptote for
$[f_{out}^{tr}(-k,t)]^*$, should be $[f_{in}^{tr}(-k,t)]^*$, rather
than $[g_1(-k,t)]^*$.

\section {Scattering time} \label{a5}

As was said above the delay times serve as an indicator of whether
the potential investigated is repulsive or attractive. Being
accumulated in the course of the scattering process, the time delays
however say little about its duration. Therefore we have to define
once more characteristic time, which will be further referred to as the
low bound of a total scattering time.

It is obvious that this quantity cannot be defined individually for
each scattering channel, because it should describe just the very stage
of the one-dimensional scattering when a particle cannot be identified
as incident, transmitted or reflected one.  Besides, in this case one
should keep in mind that the scattering event lasts until the
probability to find a particle passing through the barrier region is
noticeable. The scattering time does not coincide with the time of
staying the particle in the barrier region itself. One can say that in
order to define the scattering time, one needs to find the temporal
interval such that beyond this time gap the particle's state evolves in
the vicinity of the asymptotes.

Let us consider the left-side case and find such instant of time,
$t_{start}$, at which the distance between the CM of the incident
packet and the left boundary of the barrier is equal to the half-width
of this packet (see Appendix), i.e.,

\begin{equation} \label{300}
(a-<\hat{x}>_{in}(t_{start}))^2=<(\delta \hat{x})^2>_{in}(t_{start}).
\end{equation}
We will name the instant $t_{start}$ as the time of the onset of the
scattering event. Before this instant the state of a particle evolves
in the vicinity of the total in asymptote.

As regards the end of the scattering event, one has to take into
account that the transmitted and reflected particles move, on this
stage, in the disjoint spatial regions, i.e., their wave packets do not
interfere with each other. These packets leave the barrier at the
different instants of time. Let $t^{(1)}_{end}$ and $t^{(2)}_{end}$ be
such instants of time that
\begin{equation} \label{301}
L^2_{tr}(t^{(1)}_{end})=<(\delta \hat{x})^2>_{tr}(t^{(1)}_{end}),
\end{equation}
\begin{equation} \label{302}
L^2_{ref}(t^{(2)}_{end})=<(\delta \hat{x})^2>_{ref}(t^{(2)}_{end})
\end{equation}
(see expressions (\ref{2170}) and (\ref{2180}).

Each of equations (\ref{300}) - (\ref{302}) have two roots. A
simple analysis shows that in the case of (\ref{300}) one has to take
the smallest root. But in the case of (\ref{301}) and (\ref{302}) only
the biggest root has a physical sense. So, the searched-for solutions
to (\ref{300}) - (\ref{301}) can be written in the form

\begin{equation} \label{303}
t_{start}=\frac{m}{\hbar}\cdot\frac{a k_0-\sqrt{l^2_0k_0^2
+(a^2-l_0^2)<(\delta k)^2>_{in}}}{k_0^2 - <(\delta k)^2>_{in}}
\end{equation}
(remind that $a\gg l_0$); for $n=1,2$

\begin{eqnarray} \label{3041}
t^{(n)}_{end}=\frac{m}{\hbar \Big(k_{n}^2
-\overline{\delta k}^2_{n}\Big)} \Big(\bar{b}_{n} k_{n}-\chi_{n}
+\nonumber \\
\sqrt{\sigma_{n}k_{n}^2+\chi_{n}^2-2k_{n}\bar{b}_{n}\chi_{n}+
(\bar{b}_{n}^2-\sigma_{n})\overline{\delta k}^2_{n}}\Big);
\end{eqnarray}

\[k_{1,2}=<k>_{(tr,ref)}$,\ppp $\overline{\delta k}_{1,2} =
\sqrt{<(\delta k)^2>_{(tr,ref)}}\]
(see also expressions (\ref{2170}) and (\ref{222}) for transmission, and
(\ref{2180}) and (\ref{223}) for reflection).

The maximal time, $t^{(1)}_{end}$ or $t^{(2)}_{end}$, we take as the
end time, $t_{end},$ of scattering. Then the total scattering time,
$\tau_{scatt},$ can be defined as the difference $t_{end}-t_{start}$.

A simple analysis shows that this quantity is strictly positive when
the inequalities

\begin{equation} \label{304}
k_0> \sqrt{<(\delta k)^2>_{in}},
\end{equation}
and
\begin{equation} \label{305}
k_{n}>\overline{\delta k}_{n};\ppp n=1,2
\end{equation}
are fulfilled. They should be considered as the conditions of a
completed scattering. In this case, with the sufficient accuracy, one
can say that all incident particles start at $t=0$ toward the barrier,
and the transmitted and reflected packets occupy, in the limit
$t\to\infty$, disjoint spatial regions.  For a completed scattering the
hierarchy $t_{end}>t_{start}>0$ is obvious to be true. The second
inequality is satisfied due to (\ref{304}). The first one is valid,
since the quantum ensemble of particles cannot leave the barrier region
before entering it: the number of particles in the whole quantum
ensemble is constant.  In this case it is important to note that, in
the limit $k_0^2 \to <(\delta k)^2>_{in}$,
\[t_{start}=\frac{m(a^2-l_0^2)}{2\hbar k_0 a}\approx \frac{m a}{2\hbar
k_0}.\] That is, expression (\ref{303}) has no singularity in this
limit.

One can easily show that there is an optimal value of $l_0$ at which
$\tau_{scatt}$ is minimal: in the limit $l_0\to \infty$, the scattering
time grows together with $l_0$, but at small values of this parameter
this time is large because of the fast spreading of the wave packet. If
requirements (\ref {305}) are violated, the transmitted and reflected
packets must be overlapped at $t\to\infty$ due to their spreading. As a
result, the scattering event becomes incomplete.

In the limit $l_0\to\infty$, i.e., for narrow (in $k$-space) wave
packets, expressions (\ref{303}) - (\ref{3041}) are essentially
simplified. In this case $k_2=k_1=k_0$ and terms with $<(\delta k)^2>$,
$<(\delta k)(\delta J^\prime)>$ and $<(\delta k)(\delta F^\prime)>$ may
be neglected. Taking into account only the dominant terms in
(\ref{303}) - (\ref{3041}), we obtain

\[t_{start}=\frac{m}{\hbar k_0}(a-l_0);\]
\begin{equation} \label{3201}
t_1-t_{start}=\frac{m}{\hbar k_0} l_{scat}^{(1)},\ppp
t_2-t_{start}=\frac{m}{\hbar k_0} l_{scat}^{(2)}
\end{equation}
where
\begin{equation} \label{3202}
l_{scat}^{(1)}=l_0+<J^\prime>_{tr}+\sqrt{\sigma_1}
\end{equation}
\begin{equation} \label{3203}
l_{scat}^{(2)}=l_0+<J^\prime- F^\prime>_{ref}+\sqrt{\sigma_2}.
\end{equation}

The maximum of these differences should be taken as the total
scattering time $\tau_{scatt}$ and the corresponding quantity,
$l_{scat}^{(1)}$ or $l_{scat}^{(2)}$, may be treated, in this
particular case, as the scattering length.  Note, for the right-side
case the sign of $F^\prime(k)$, in the expressions for scattering time
and length, is opposite. For symmetric potential barriers this
derivative is equal to zero for all $k$.

\section*{Conclusion}

One of the main aims of this paper was to argue that in the case of a
one-dimensional completed scattering the conventional quantum theory
quite admits a separate description of transmitted and reflected
particles at the stage preceding the scattering event. We introduced
the notion of the transmission and reflection channels and showed that
for each channel one can define an unique (unitary) scattering matrix.
By the known out asymptotes describing separately transmitted and
reflected particles, these matrices enable one to reconstruct uniquely
in asymptotes to describe separately to-be-transmitted and
to-be-reflected particles. As was shown, the momentum distribution and
the number of particles calculated over the in and out asymptotes, for
each channel, should be the same. Besides, in both scattering channels
particles were shown to start, on the average, from the same point. As
regards the second and higher moments of the position operator, they
cannot be determined separately for these two subensembles of
particles.

On this basis we proposed characteristic times to describe tunneling a
spinless particle through a time-independent potential barrier. We
introduced the individual delay times for transmission and reflection.
Besides, to estimate the duration of the scattering event, we derived
the expression for the (total) scattering time. All three tunneling
times can be applied for wave packets of an arbitrary width. They were
obtained in terms of the mean values of the particle's momentum and
phase times calculated over the incident, transmitted or reflected
wave packets. In addition, we found the conditions to be satisfied for a
completed scattering.

\appendix

\section*{Expectation values of operators over in  and out-states}

Let us consider the left-side case (see Exp. (\ref{88}) and calculate
all needed moments of the position and momentum operators over the
incident, transmitted and reflected wave packets. In the limit $t\to
-\infty$ we have to deal with the incident one to move along
in asymptote (\ref{59}). In the limit $t\to \infty$ we have to consider
transmitted and reflected packets to correspond to separate
out asymptotes (\ref{61}) and (\ref{62}) (separate in asymptotes
(\ref{590}) and (\ref{610}) can be treated by the same manner). Let us
write down these wave functions in the form

\newcommand {\intk}{\int_{-\infty}^{\infty}dk}
\newcommand {\intx}{\int_{-\infty}^{\infty}dx}
\newcommand {\da}{\partial}

\begin{equation} \label{201}
\Psi(x,t)=\frac{1}{\sqrt{2\pi}} \intk f(k,t) e^{ikx},
\end{equation}
\[f(k,t)=M\aro \exp(i\xi(k,t))\]
(remind that $f(k,t)\in {\cal S}_{\infty}$);
$M\aro$ and $\xi(k,t)$ are the real functions. In particular, for the
incident packet (\ref{59}) we have

\begin{equation} \label{202}
M_{in}\aro=|\Api(k)|; \ppp
\xi_{in}(k,t)=-\frac{\hbar k^2 t}{2m}.
\end{equation}
For the transmitted (\ref{61}) and reflected (\ref{62}) packets,
\begin{eqnarray} \label{203}
M_{out}^{tr}\aro=\sqrt{T(k)}M_{in}\aro; \nonumber\\
\xi_{out}^{tr}(k,t)=\xi_{in}(k,t)+J(k)-kd;
\end{eqnarray}
\begin{eqnarray} \label{204}
M_{out}^{ref}(-k)=\sqrt{R(k)}M_{in}\aro;  \nonumber\\
\xi_{out}^{ref}(-k,t)=\xi_{in}(k,t)+2ka+J(k)-F(k)-\frac{\pi}{2}.
\end{eqnarray}

Fourier transformation (\ref{201})-(\ref{204}) enables one to determine
the time dependence of the expectation value $<\hat{Q}>$ for any
Hermitian operator $\hat{Q}$, at the stages preceding and following the
scattering event,

\begin{equation} \label{205}
<\hat{Q}>=\frac{<\Psi|\hat{Q}|\Psi>}{<\Psi|\Psi>},
\end{equation}
where $\Psi$ is one of the above wave functions (for a subensemble of
particles this value acquires the status of a conditional probability).

Note that for the incident and reflected packets the integrals in
(\ref{205}) should be calculated, strictly speaking, over the interval
$(-\infty, a]$. For the transmitted packet, one needs to integrate over
the interval $[b,\infty)$. Expressions (\ref{201})-(\ref{204}) for
these packets are valid only for the corresponding spatial region and
corresponding stage of scattering. However, taking into account that
the body of each packet is located, in the limit $t\to \infty$ or $t\to
-\infty$, in its "own" spatial region, we may extend the integration
in (\ref{205}) onto the whole $OX$-axis. Due to this step the
description of these packets becomes very simple. As regards a mistake
introduced in the formalism, one can expect that it is sufficiently
small; in any case, the farther is the packet from the barrier at the
initial time, the smaller is this mistake. It vanishes in the limits
$t\to \mp\infty$ when the particle's state moves along the in  and
out asymptotes. Thus, the $k$-representation provides a suitable basis
for the calculation of desired characteristics of all three packets.

\subsection{Normalization}

So, within the above accuracy, the norms of these functions are
constant beyond the scattering region,

\begin{eqnarray} \label{206}
<\Psi(x,t)|\Psi(x,t)>=\intk M^2\aro.
\end{eqnarray}
Then for each packet we have the following norms. Since
$\Psi^{(0)}_{left}$ is normalized function, we have

\begin{equation} \label{207}
<\Psi^{(0)}_{left}|\Psi^{(0)}_{left}>=\intk M^2_{in}\aro=1.
\end{equation}
For the transmitted packet,
\begin{eqnarray} \label{209}
<f_{out}^{tr}|f_{out}^{tr}>=\intk [M_{out}^{tr}\aro]^2 \nonumber\\
=\intk T(k)M^2_{in}\aro \equiv <T(k)>_{in}\equiv \bar{T}.
\end{eqnarray}
For the reflected packet,
\begin{eqnarray*}
<f_{out}^{ref}|f_{out}^{ref}>=\intk [M_{out}^{ref}\aro]^2 \nonumber \\
=\intk R(k)M^2_{in}(-k).
\end{eqnarray*}
Having made an obvious change of variables, we obtain

\begin{equation} \label{210}
<f_{out}^{ref}|f_{out}^{ref}>=<R(k)>_{in}\equiv
\bar{R}.
\end{equation}
From (\ref{207}) - (\ref{210}) it follows that
\begin{equation} \label{2100}
\bar{T}+\bar{R}=1.
\end{equation}

\subsection{The expectation values of the operators $\hat{k}^n$ ($n$
is the positive number)}

Since in the $k$-representation $\hat{k}$ is a multiplication operator,
for any number $n$ we have
\begin{equation} \label{211}
<\Psi|\hat{k}^n|\Psi>=\intk M^2\aro k^n.
\end{equation}
Now we can treat separate packets. From (\ref{211}) and
(\ref{203}) it follows that
\[<f_{out}^{tr}|k^n|f_{out}^{tr}>=<f_{in}|T(k)k^n|f_{in}>.\]
In a similar way we find also that
\[<f_{out}^{ref}|k^n|f_{out}^{ref}>=(-1)^n<f_{in}|R(k)k^n|f_{in}>,\]
and, hence,
\begin{eqnarray} \label{511}
<T(k)k^n>_{in}=\bar{T}<k^n>_{tr},\nonumber\\
<R(k)k^n>_{in}=(-1)^n\bar{R}<k^n>^{ref}.
\end{eqnarray}
As a consequence, the next relationship is obvious to be valid
\begin{eqnarray} \label{212}
<k^n>_{in}=\bar{T}<k^n>_{out}^{tr}+\bar{R}<(-k)^n>_{out}^{ref}.
\end{eqnarray}

\subsection{The expectation values of the operator $\hat{x}$}

We begin again with expressions to be common for all three packets.
Since $\hat{x}=i\frac{\da}{\da k}$ we have
\begin{eqnarray}
\label{214} <\Psi|\hat{x}|\Psi>=i\intk f^*(k,t)\frac{\da f(k,t)}{\da
k}= \nonumber\\ = i\intk M\aro\frac{dM\aro}{dk} -\intk
M^2\aro\frac{\da\xi(k,t)}{\da k}.
\end{eqnarray}
Since the first term here is equal to
\[\frac{i}{2}M^2\aro|^{+\infty}_{-\infty}=0,\] we have
\begin{eqnarray} \label{215}
<\Psi|\hat{x}|\Psi>=-\intk M^2\aro \frac{\da\xi(k,t)}{\da
k} \nonumber\\ \equiv -<f|\frac{\da\xi(k,t)}{\da k}|f>.
\end{eqnarray}

For the incident and transmitted packets, taking into account
expressions (\ref{202}) and (\ref{203}) for $\xi(k,t)$,  we obtain

\begin{equation} \label{216}
<\hat{x}>_{in}= \frac{\hbar t}{m}k_0,
\end{equation}
\begin{equation} \label{217}
<\hat{x}>_{out}^{tr}= \frac{\hbar t}{m}<k>_{out}^{tr}-
<J^\prime(k)>_{out}^{tr}+d.
\end{equation}

Since the functions $J^\prime(k)$ and $F^\prime(k)$ are even, from
(\ref{204}) it follows that
\begin{equation} \label{218}
<\hat{x}>_{out}^{ref}
=2a+<J^\prime(k)-F^\prime(k)>_{out}^{ref}-\frac{\hbar
t}{m}<-k>_{out}^{ref}.
\end{equation}

Let, at the instant $t,$ $L_{tr}$ be the distance between the CM of the
transmitted packet and the nearest boundary of the barrier, i.e.,
$L_{tr}=<\hat{x}>_{tr}-b$. Similarly, let $L_{ref}$ be the distance
between the CM and the corresponding barrier's boundary for the
reflected packet at the same instant:  $L_{ref}=a-<\hat{x}>_{ref}$.
From (\ref{217}) and (\ref{218}) it follows that

\begin{equation} \label{2170}
L_{tr}(t)= \frac{\hbar t}{m}<k>_{tr}-\bar{b}_{1},
\end{equation}
\begin{equation} \label{2180}
L_{ref}(t)=\frac{\hbar t}{m}<-k>_{ref}-\bar{b}_{2}
\end{equation}
where $\bar{b}_{1}=<J^\prime(k)>_{out}^{tr}+a,$
$\bar{b}_{2}=<J^\prime(k)-F^\prime(k)>_{out}^{ref}+a.$

\subsection{The mean-square deviations in $x$-space}

Let us derive firstly the expression for all packets. We have
\[<\Psi|\hat{x}^2|\Psi>=-\intk f^*(k,t)\frac{\da^2f(k,t)}{\da
k^2}.
\]
Since
\[\frac{\da^2f(k,t)}{\da k^2}=\left[M''-M(\xi')^2+i(2M'\xi'+M\xi'')
\right]e^{i\xi},\]
we have
\begin{eqnarray} \label{219}
<\Psi|\hat{x}^2|\Psi>=\intk M [M(\xi')^2-M'']\nonumber\\ -i\intk
[(M^2)'\xi'+M^2\xi'']
\end{eqnarray}
One can easily show that the last integral in (\ref{219}) is equal to
zero. Therefore

\begin{eqnarray} \label{220}
<\Psi|\hat{x}^2|\Psi>=\intk M^2\aro
[\xi'(k,t)]^2 \nonumber \\+\intk [M'\aro]^2.
\end{eqnarray}

Let, for any operator $\hat{Q}$, $<(\delta \hat{Q})^2>$ be the
mean-square deviation: $\delta\hat{Q} =\hat{Q}-<\hat{Q}>$. For the
operator $\hat{x}$ we have

\begin{equation} \label{400}
<(\delta \hat{x})^2>=<(\ln'M)^2>+<(\delta\xi')^2>.
\end{equation}

Now we are ready to find these quantities for each packet.
Using (\ref{400}) and expressions (\ref{202})-(\ref{204}), one can show
that for incident packet
\begin{equation} \label{221}
<(\delta \hat{x})^2>_{in}=<(\ln^\prime A)^2>_{in}+
\frac{\hbar^2t^2}{m^2} <(\delta k)^2>_{in}
\end{equation}
(here the first term is equal to $l_0^2,$ in accordance with the
initial condition); for the transmitted packet
\begin{equation} \label{222}
<(\delta \hat{x})^2>_{out}^{tr}= \sigma_{1}-
2\frac{\hbar t}{m}\chi_{1}+
\frac{\hbar^2t^2}{m^2} <(\delta k)^2>_{out}^{tr};
\end{equation}
for the reflected packet
\begin{eqnarray} \label{223}
<(\delta \hat{x})^2>_{out}^{ref} =\sigma_{2} -2\frac{\hbar
t}{m}\chi_{2}\nonumber \\ +\frac{\hbar^2t^2}{m^2} <(\delta
k)^2>_{out}^{ref};
\end{eqnarray}
here
\[\sigma_{1}= <(\ln^\prime M_{out}^{tr})^2>_{out}^{tr}+<(\delta
J^\prime)^2>_{out}^{tr};\] \[\sigma_{2}=<(\ln^\prime
M_{out}^{ref})^2>_{out}^{ref}+<(\delta J^\prime -\delta
F^\prime)^2>_{out}^{ref};\]
\[\chi_{1}=<(\delta J^\prime)(\delta k)>_{out}^{tr},\hspace{2mm}
\chi_{2}=-<(\delta J^\prime-\delta F^\prime)(\delta k)>_{out}^{ref}\]

\end{document}